\documentclass[11pt]{amsart}
\usepackage{geometry}                
\usepackage{graphicx}
\usepackage{amssymb}
\usepackage{epstopdf}
\usepackage{array}
\usepackage{subfigure}
\usepackage{url}
\usepackage{hyperref}
\usepackage{xspace}
\usepackage{lscape}
\usepackage{isorot}

\usepackage{color}

\definecolor{gray}{rgb}{0.7,0.7,0.7}

\usepackage{supertabular}
\usepackage{hhline}
\newcommand\arraybslash{\let\\\@arraycr}
\makeatother
\title[Slimy whiskers]{Slimy hairs: \\ Hair sensors made with slime mould}
\author[Adamatzky]{ 
 Andrew Adamatzky \vspace{0.5cm}\\  {\footnotesize{University of the West of England, Bristol, UK \\ andrew.adamatzky@uwe.ac.uk}}}
\address[Adamatzky]{University of the West of England, Bristol, United Kingdom}

\begin{document}

\maketitle

\begin{abstract}
Slime mould \emph{Physarum polycephalum} is a large single cell visible by unaided eye.  We design a slime mould implementation of a tactile hair, 
where the slime mould responds to repeated deflection of hair by an immediate high-amplitude spike and a prolonged increase in amplitude and width of its oscillation  impulses. We demonstrate that signal-to-noise ratio of the Physarum tactile hair sensor averages near 6 for the immediate response and 2 for the prolonged response.

\end{abstract}

\section{Introduction}

Tactile sensors are ubiquitous in robotics and medical devices~\cite{Cutkosky_2008, Hamed_2008, Mukai_2008, Lucarotti_2013} and thus 
development of novel types of these sensing devices remains a hot topic of engineering, material sciences and bionics.   Novel designs and implementations, see  overviews in~\cite{Rocha_2008,Lucarotti_2013}, include arrays of electro-active polymers and ionic polymer metal composites~\cite{wang_2008, wang_2009}, piezoelectric polymer oxide semiconductor field effect transistors tactile arrays~\cite{Dahiya_2009}, 
pressure sensitive conductive rubber~\cite{Kato_2008, Ohmukai_2012}, flexible capacitive micro-fluidic based sensors~\cite{Wong_2012}, 
and patterns of micro-channels filled with eutectic gallium-indium~\cite{Park_2010, Park_2011}. Recently an interest in technological 
developments started to move away from solid materials to soft matter implementations, see overview in~\cite{Tiwana_2012}, and 
bio-inspired and hybrid implementations: bio-mimetic sensors which employ a conductive fluid encapsulated in elastic container and use
deformation of the elastic container in transduction~\cite{Wettels_2008}, carbon nanotube filled elastomers~\cite{Engel_2006}, 
polymer hair cell sensors~\cite{Engel_2006a}. 

Live cell sensors~\cite{Taniguchi_2010} and bio-hybrid sensors encapsulating living fibroblasts as a part of transduction system~\cite{Cheneler_2012} are of particular interest because they open totally new dimension in engineering sensing technologies. The living substrates used in sensor 
design are too dependent on a 'life-support' system, they need supply of nutrients and removal of waste. Therefore we decided to consider
an autonomous living creature which does not require sophisticate support bandwidth and can survive for a long period of time without 
a laboratory equipment. This is a vegetative stage, a plasmodium of acellular slime mould \emph{Physarum polycephalum}.

\begin{figure}[!tbp]
\centering
\includegraphics[width=0.7\textwidth]{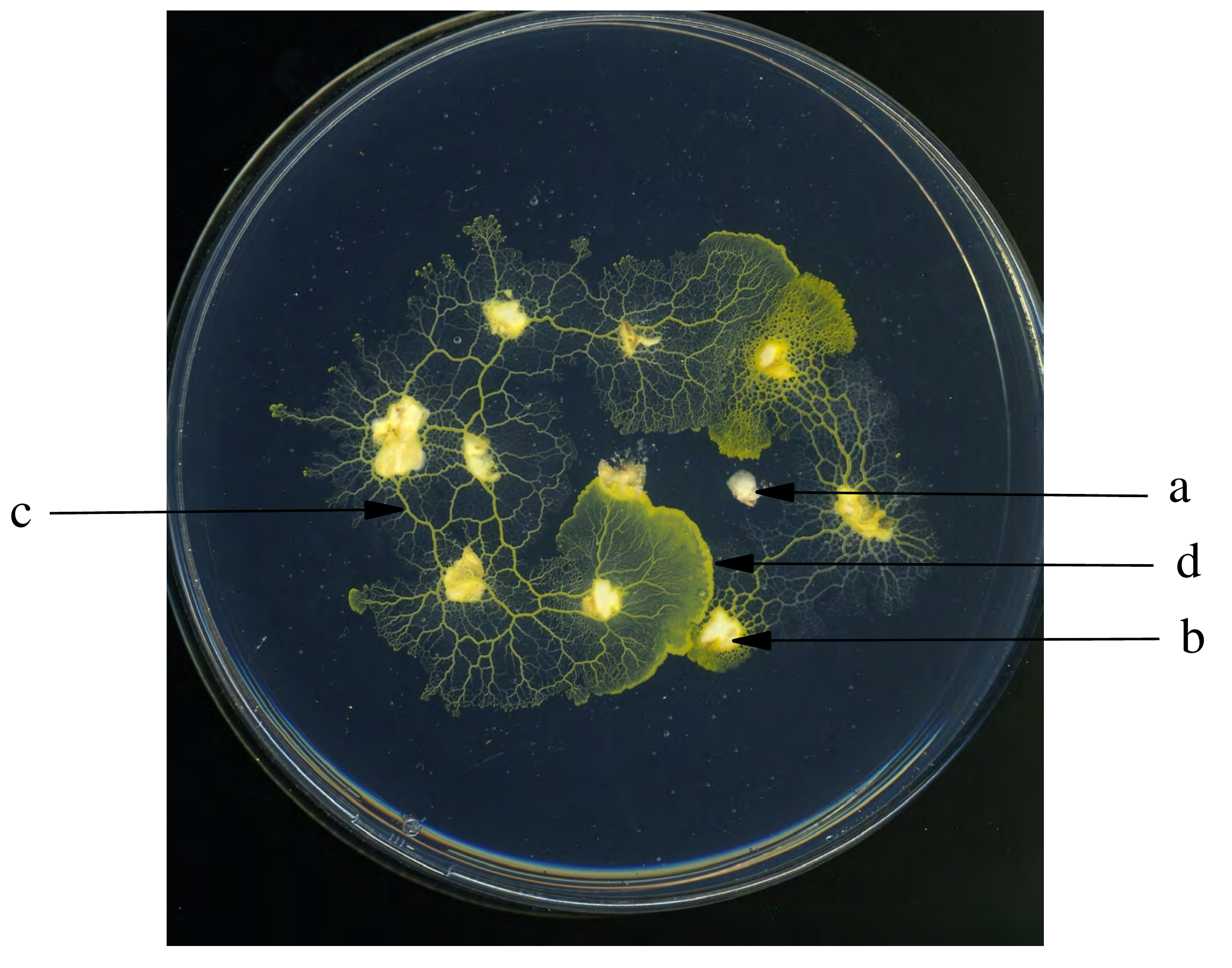}
\caption{Plasmodium of \emph{P. polycephalum} on a data set, planar configuration of oat flakes, on an agar gel in a Petri dish 8~cm diameter.
(a)~A virgin oat flake.
 (b)~An oat flake colonised by the plasmodium.
 (c)~A protoplasmic tubes.
 (d)~An active zones, growing parts of the plasmodium. 
}
\label{PhysExm}
\end{figure}

\emph{Physarum polycephalum} belongs to the species of order \emph{Physarales}, subclass \emph{Myxogastromycetidae}, class \emph{Myxomycetes}, division \emph{Myxostelida}. It is commonly known as a true, acellular or multi-headed slime mould. Plasmodium is a single 
cell with a myriad of diploid nuclei.  The plasmodium is visible to the naked eye (Fig.~\ref{PhysExm}). The plasmodium looks like an amorphous yellowish mass with networks of protoplasmic tubes. The plasmodium behaves and moves as a giant amoeba. It feeds on bacteria, spores and other microbial creatures and micro-particles~\cite{stephenson_2000}. 

Plasmodium of \emph{P. polycephalum} consumes microscopic particles, and during its  foraging behaviour the plasmodium spans scattered sources of nutrients with a network of  protoplasmic tubes (Fig.~\ref{PhysExm}).  The plasmodium optimises its protoplasmic network that covers all sources of nutrients and  guarantees robust and quick distribution of nutrients in the plasmodium's body. Plasmodium's foraging behaviour can be  interpreted as a computation:  data are represented by spatial distribution of attractants and repellents, and  results are represented by a structure of protoplasmic  network~\cite{adamatzky_physarummachines}.  Plasmodium can solve computational problems with natural parallelism, e.g. related to shortest 
path~\cite{nakagaki_2000} and hierarchies of planar proximity graphs~\cite{adamatzky_ppl_2008}, computation of plane 
tessellations~\cite{shirakawa}, execution of logical computing schemes~\cite{tsuda2004,adamatzky_gates}, and natural implementation of spatial logic and process algebra~\cite{schumann_adamatzky_2009}.  In~\cite{adamatzky_physarummachines} we experimentally demonstrated that slime mould  \emph{P. polycephalum} is a programmable amorphous biological computing substrate, capable for solving a wide range of tasks: from computational geometry and  optimisation on graphs to logics and general purpose computing. 

In our previous paper on slime mould's tactile sensors~\cite{adamatzky_2013_tactile} we experimentally studied how Physarum reacts to 
application of a load to its fine network of tubes or a single tube. We demonstrated impulse and pattern of oscillation responses and characterised
sensorial abilities of Physarum. We found, however, that it is not practical to allow Physarum tactile sensor to be in a direct contact with a load, because
the slime mould starts colonising the load and becomes attached to the load and the sensor could be damaged when the load is lifted. Thus 
some intermediary mechanical medium is required an object causing tactile stimulation and the Physarum. Looking for alternative solutions we encountered Engel and colleagues design~\cite{Engel_2006a} of a spider artificial hair made of polyurethane and fixed to a flexible 
substrate.  Being inspired by their approach we implemented a slime mould-based analog of a spider tactile hair outlined in 
present paper.

\section{Methods}
\label{methods}

\begin{figure}[!tbp]
\centering
\includegraphics[width=0.8\textwidth]{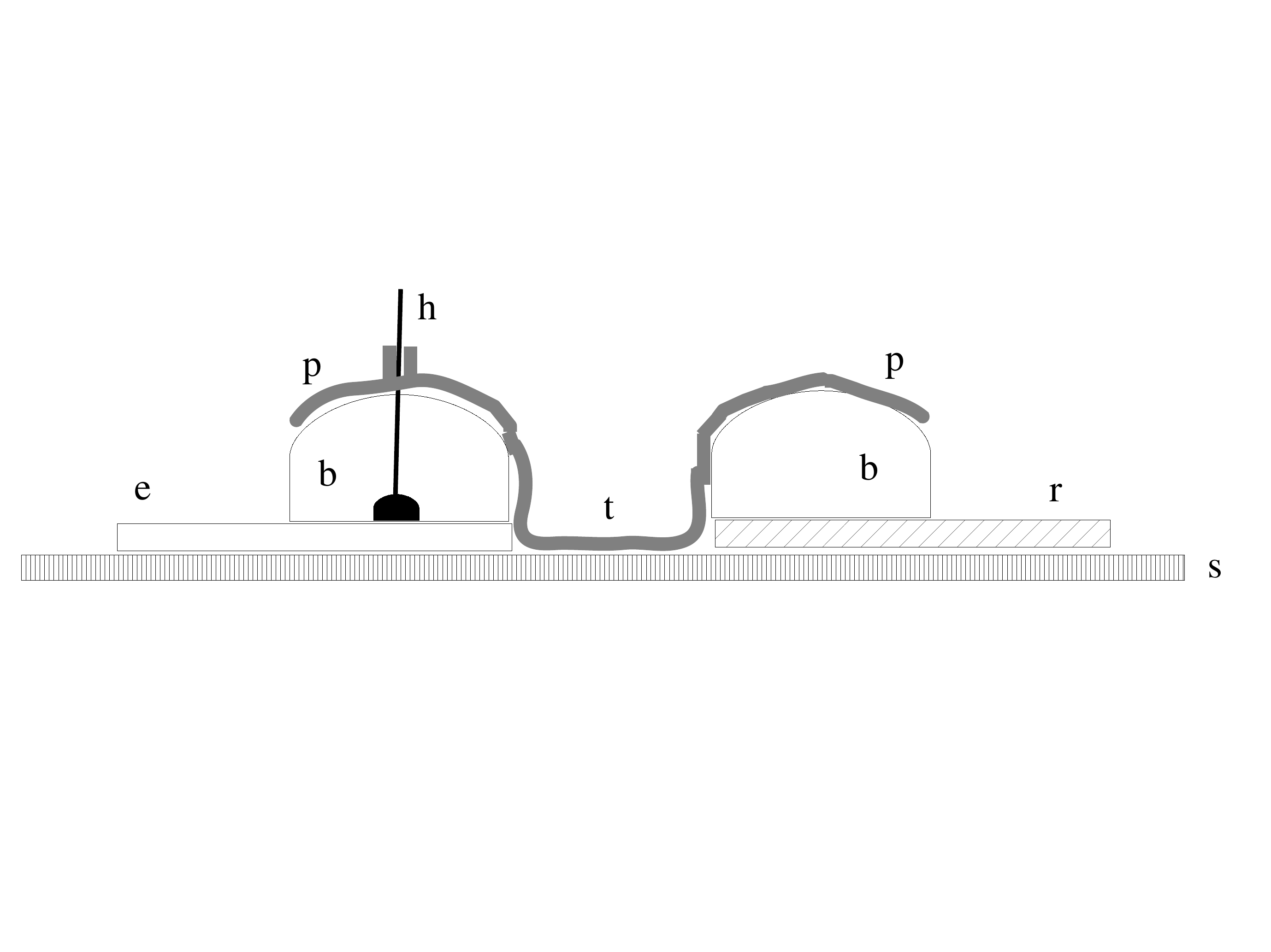}
\caption{Scheme of experimental setup. Electrodes (r) and (e) are fixed to a bottom of polyurethane Petri dish.
A hair (h) is fixed to one electrode with epoxy glue. Agar blobs (b) are placed on electrodes. The hair penetrates 
through the centre of one agar blob. Physarum (p) colonises both agar blobs, propagates across bare plastic substrate (s) 
and connects two blobs with a protoplasmic tube (t). Root of the hair is partly colonised by Physarum.}
\label{scheme}
\end{figure}

Plasmodium of \emph{Physarum polycephalum} was cultivated in plastic lunch boxes (with few holes punched in their lids for ventilation) on wet kitchen towels and fed with oat flakes. Culture was periodically replanted to fresh substrate. A scheme of experimental setup is shown in Fig.~\ref{scheme}. A planar aluminium foil electrodes (width 5~mm, 0.04~mm thick, volume resistance 0.008~$\Omega$/cm$^2$) are  fixed  to a bottom of a plastic Petri dish (9~cm), see Fig.~\ref{scheme}er, where 'r' is a reference electrode and 'e' is recording electrode. Distance between proximal sites of electrodes is 10~mm. One or more hairs (either human hairs or bristles from a toothbrush), 10 mm lengths, were fixed by upright to electrode 'e' using epoxy glue. Then 2~ml of agar was gently and slowly powered onto the electrodes to make a dome like blobs of agar (Fig.~\ref{scheme}b). An oat flake occupied by Physarum was placed on an agar blob, residing on a reference electrode 'r'. Another oat flake not colonised by Physarum was placed on an agar blob on electrode 'e'.  Physarum exhibits chemotaxis behaviour and therefore propagates on a bare bottom of a Petri dish from blob 'r' to blob 'e', usually in 1-3 days. Thus in 1-3 days after inoculation of Physarum on blob 'r', both blobs became colonised by Physarum (Fig.~\ref{scheme}p), and the blobs  became connected by a single protoplasmic tube  (Fig.~\ref{scheme}t). Experiments where more than one tube connected blobs with Physarum develop were usually discounted because patterns of oscillation were affected by interactions between potential waves travelling along interlinked protoplasmic tubes.  Electrical activity of plasmodium was recorded with  ADC-24 High Resolution Data Logger  (Pico Technology, UK), a recording is taken every 10~sec (as much samples as possible are averaged during this time window). Tactile stimulation was provided by deviating whisker from its original position 30 times, two times per second; a tip of the whisker was deflected from its position by 20-30 degrees. Whiskers were deflected using a 15~cm insulator stick.

\begin{figure}[!tbp]
\centering
\includegraphics[width=0.6\textwidth]{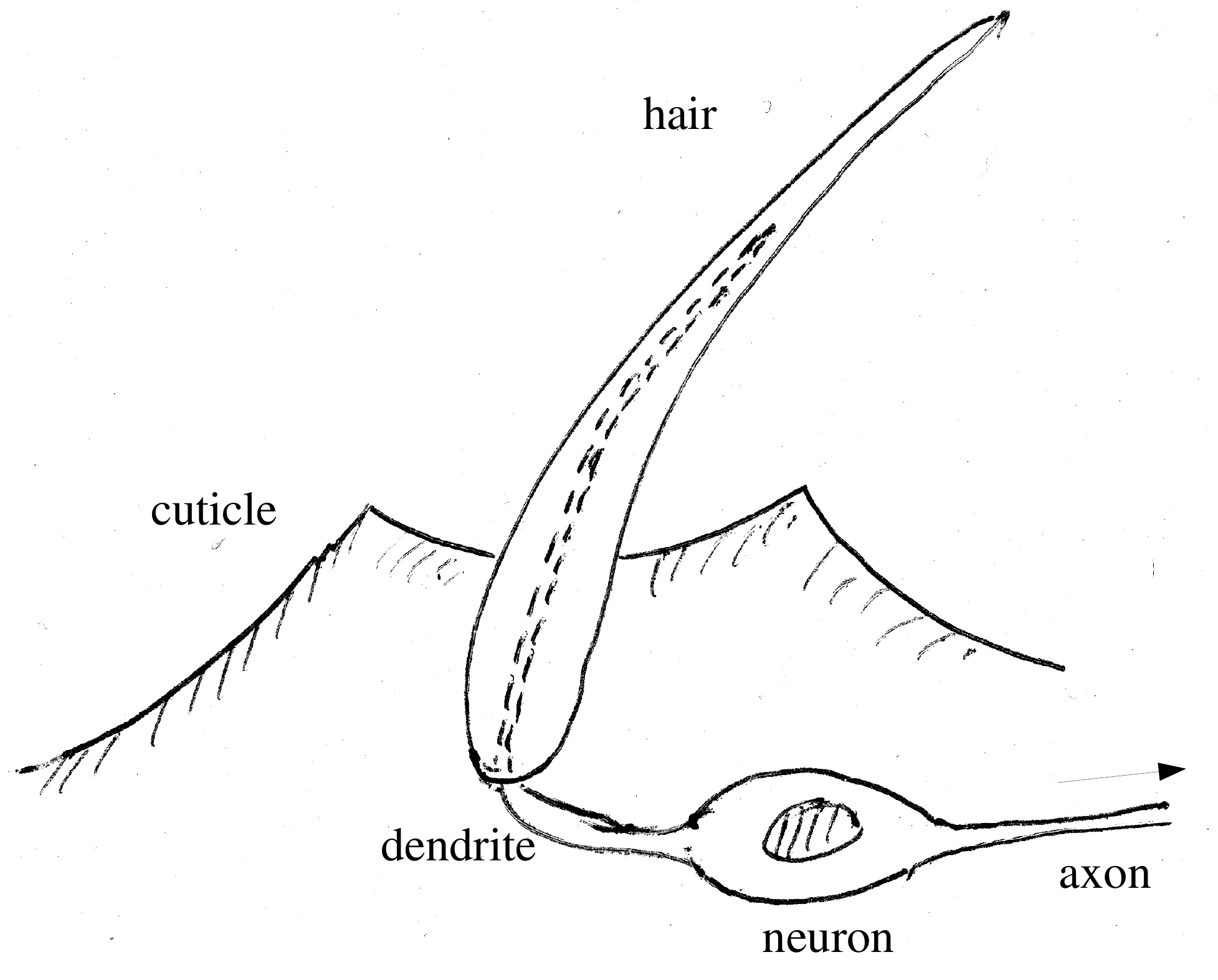}
\caption{A scheme of a spider tactile hair. Adapted from~\cite{barth_2004}.}
\label{schemehair}
\end{figure}

In our experimental setup we aimed to imitate, at least at an abstract level,  a spider tactile hair, see Fig.~\ref{schemehair} and details in~\cite{barth_2004}. This is a medium size seta, or hair, with slightly curved shaft. This seta is usually innervated by 1-3 neurons, one neuron is shown in 
Fig.~\ref{schemehair}, which dendrites enter the shaft but may not propagate till top of the hair. We imitated seta with a natural or synthetic 
bristle without shaft (Fig.~\ref{scheme}h), rightly predicting that slime mould will climb on the bristles. Dendrite is imitated by plasmodial network 
on electrode 'e', see Fig.~\ref{scheme}; neuron is a protoplasmic tube (Fig.~\ref{scheme}t) connecting the agar blobs, and axon is a plasmodial network  on electrode 'r'.

\section{Results}
\label{results}

\begin{figure}[!tbp]
\centering
\subfigure[]{\includegraphics[width=0.5\textwidth]{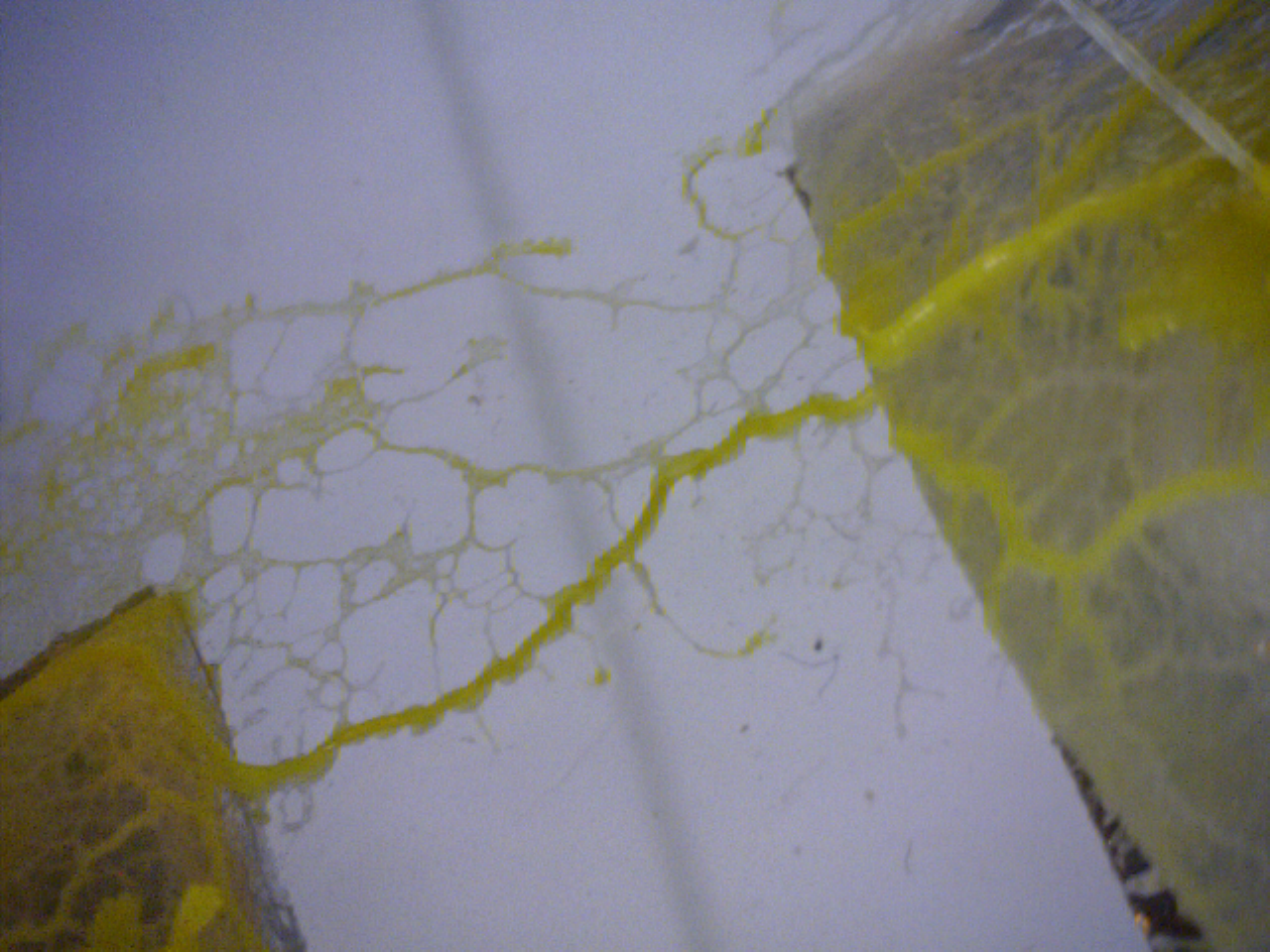}}
\subfigure[]{\includegraphics[width=0.35\textwidth]{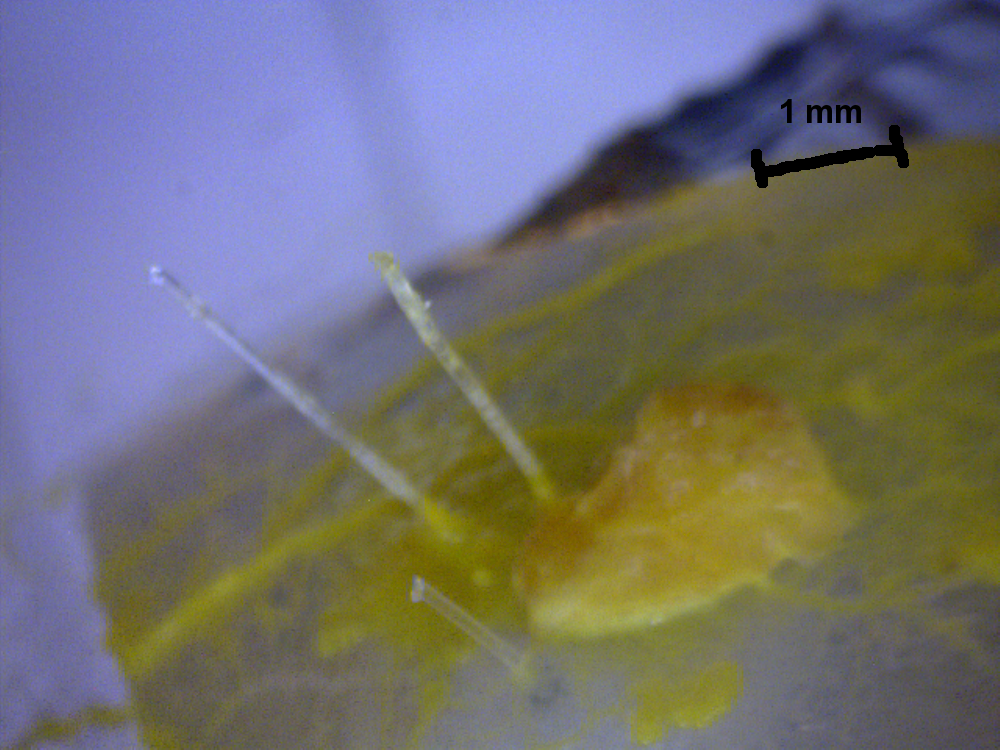}}
\subfigure[]{\includegraphics[width=0.45\textwidth]{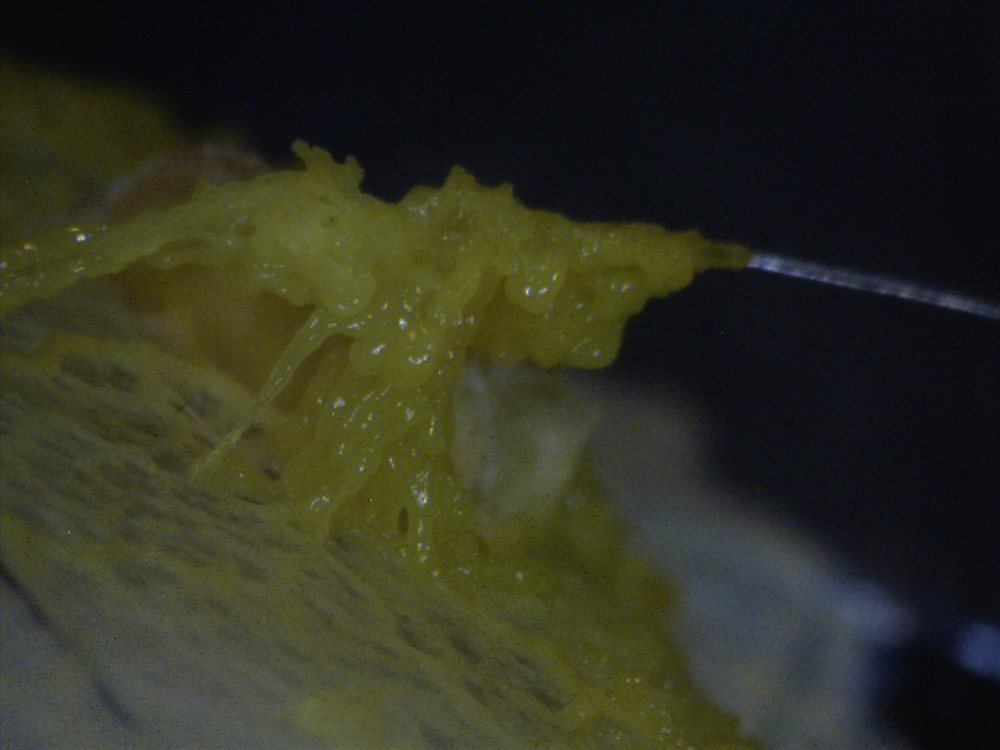}}
\subfigure[]{\includegraphics[width=0.25\textwidth]{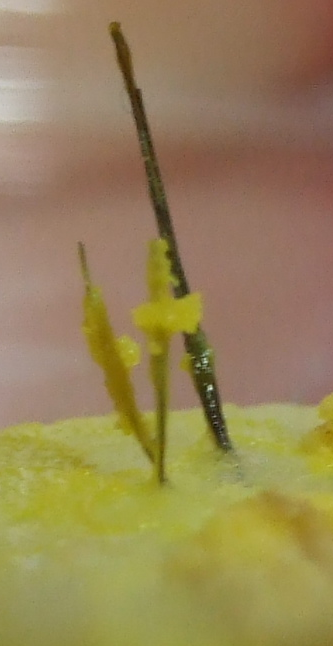}}
\caption{Photographs of Physarum setup. (a)~A protoplasmic tube connecting blobs of agar. (bcd)~Whiskers partly or completely occupied by Physarum.}
\label{photowhiskers}
\end{figure}

In 1-3 days after inoculation of Physarum to an agar blob on a references electrode it propagates to and colonises agar blob 
on a recording electrode.  A living wire --- protoplasmic tube --- connecting two blobs of Physarum is formed (Fig.~\ref{photowhiskers}a).  
In most cases Physarum 'climbs' onto hairs/bristles and occupies one third to a half of their length  (Fig.~\ref{photowhiskers}b). In many 
cases a sub-network of protoplasmic tubes is formed around the base of the hair/bristle (Fig.~\ref{photowhiskers}c). In some cases 
Physarum  occupies the whole hair/bristle, from the bottom to the top (Fig.~\ref{photowhiskers}d). 

\begin{figure}[!tbp]
\centering
\subfigure[]{
\includegraphics[width=0.6\textwidth]{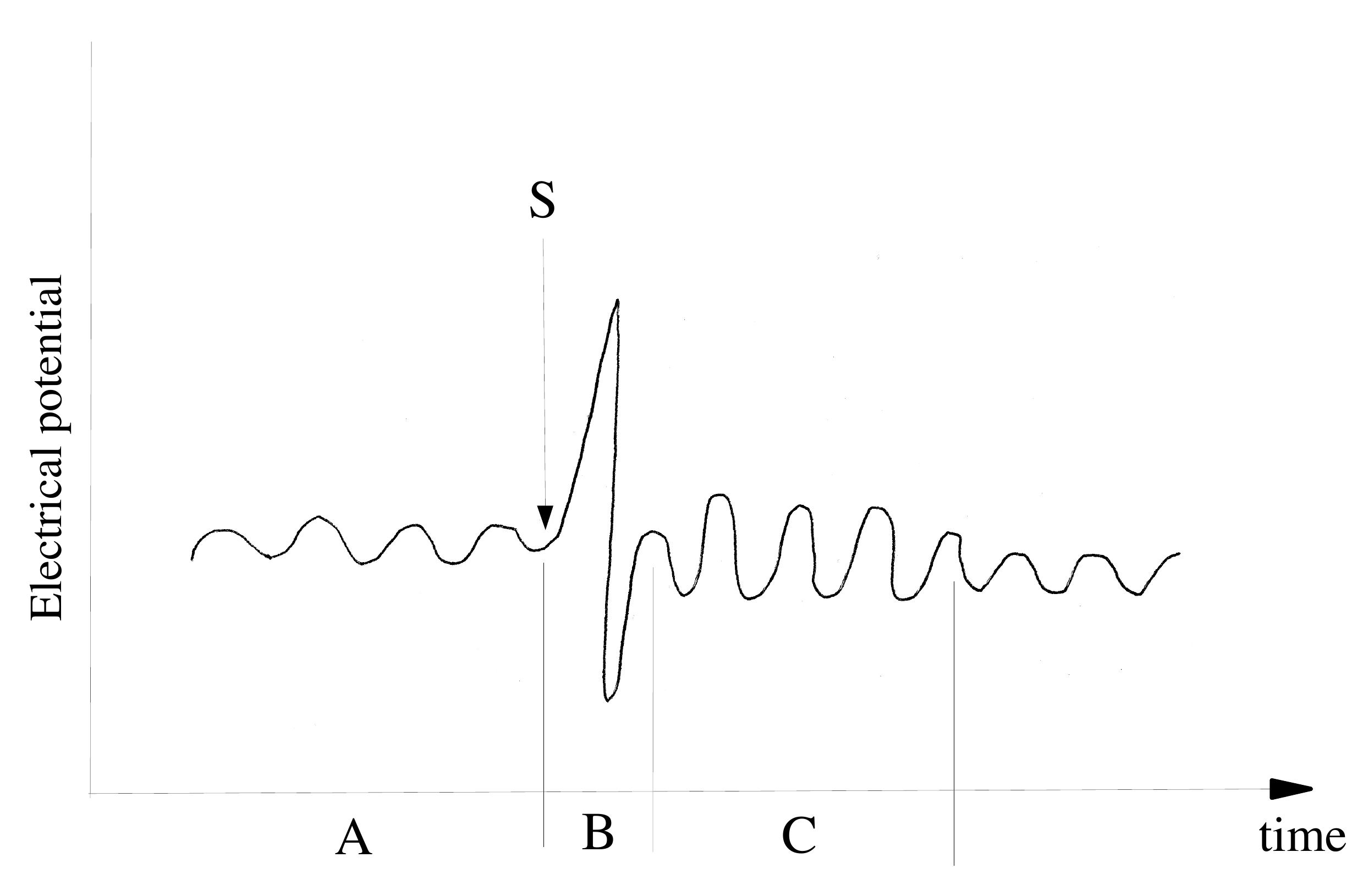}
}
\subfigure[]{
\includegraphics[width=0.6\textwidth]{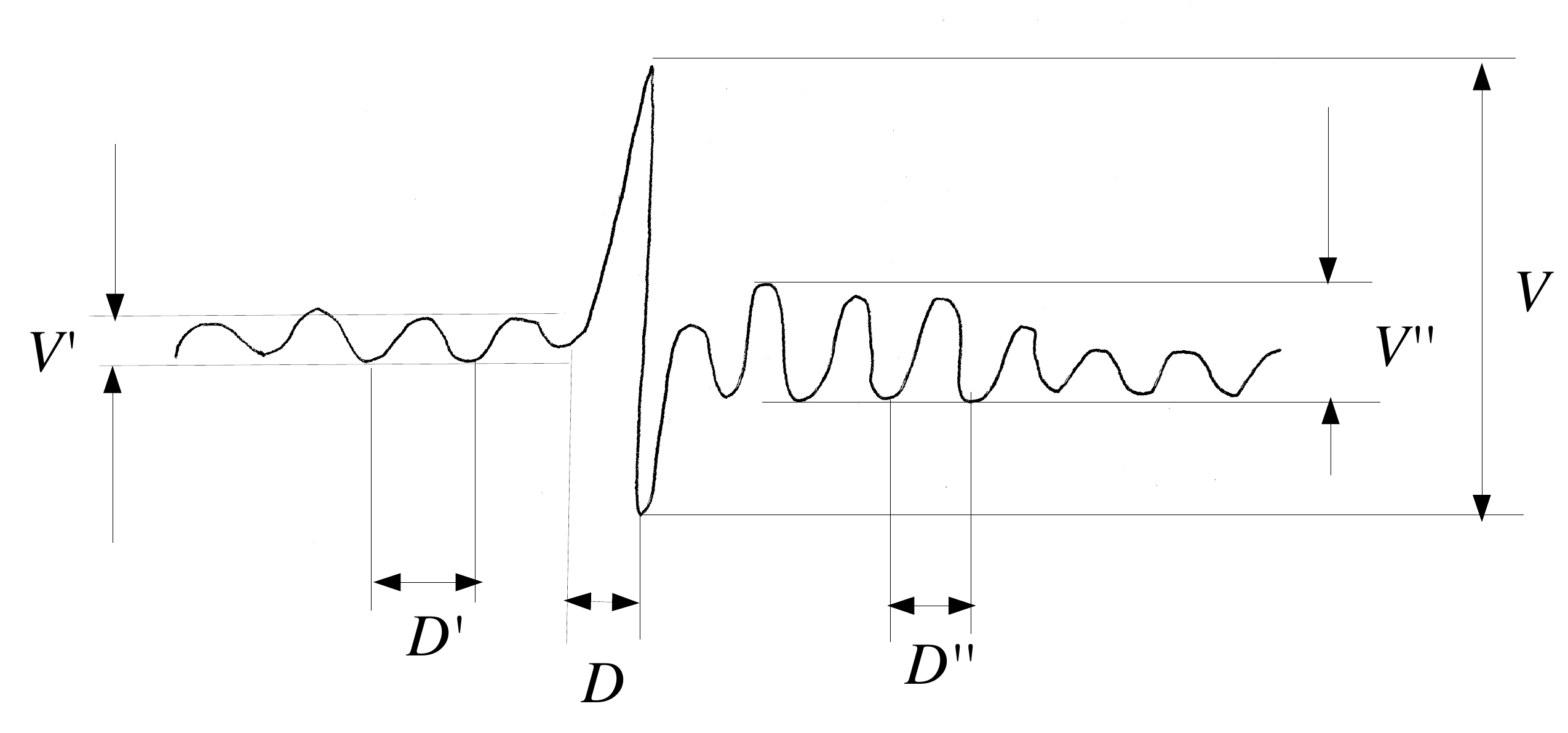}
}
\caption{
(a)~Stages of electrical potential dynamics: period 'A' is a normal oscillator activity, 'S' is a moment of whisker repeated deviation, 'B' is an immediate response, high-amplitude spike, 'C' is a prolonged response, an envelop of waves. 
(b)~Scheme of parameters measured: $V$ is an amplitude  of a high-amplitude response spike, $V'$ and $V''$ are amplitudes of oscillations before and after stimulation, averaged by closest oscillations; $D$ is a width of a high-amplitude response spike, and $D'$ and $D''$ are average widths of oscillations before and after stimulation. Amplitudes are measured in mV and widths in seconds.}
\label{parameters}
\end{figure}

An undisturbed Physarum exhibit more or less regular patterns of oscillations of its surface electrical potential (Fig.~\ref{parameters}a, A). 
The electrical potential oscillations are more likely controlling a peristaltic activity of protoplasmic tubes, necessary for distribution of nutrients in the spatially extended body of Physarum~\cite{seifriz_1937,heilbrunn_1939}. A calcium ion flux through membrane triggers oscillators 
responsible for dynamic of contractile activity~\cite{meyer_1979,fingerle_1982}.  It is commonly acceptable is that the potential oscillates with amplitude of 1 to 10~mV and period 50-200~sec, associated with shuttle streaming of cytoplasm~\cite{iwamura_1949, kamiya_1950, kashimoto_1958, meyer_1979}.  In our experiments we observed sometimes lower amplitudes because there are agar blobs between Physarum and electrodes and, also, electrodes were connected with protoplasmic tube only. Exact characteristics of electric potential oscillations vary depending on 
state of Physarum culture and experimental setups~\cite{achenbach_1980}.

\begin{figure}[!tbp]
\centering
\subfigure[]{\includegraphics[width=0.6\textwidth]{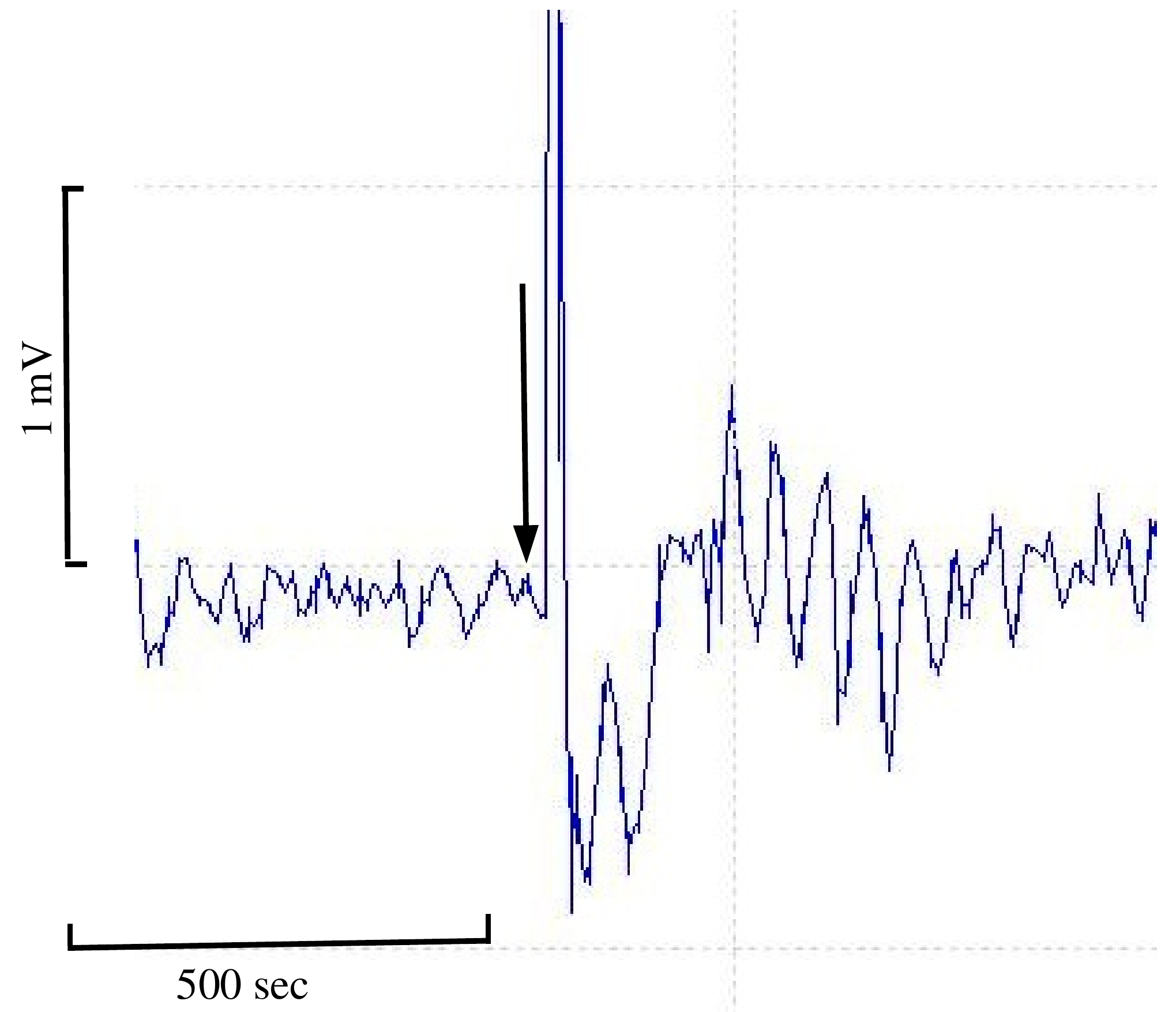}}
\subfigure[]{\includegraphics[width=1.1\textwidth]{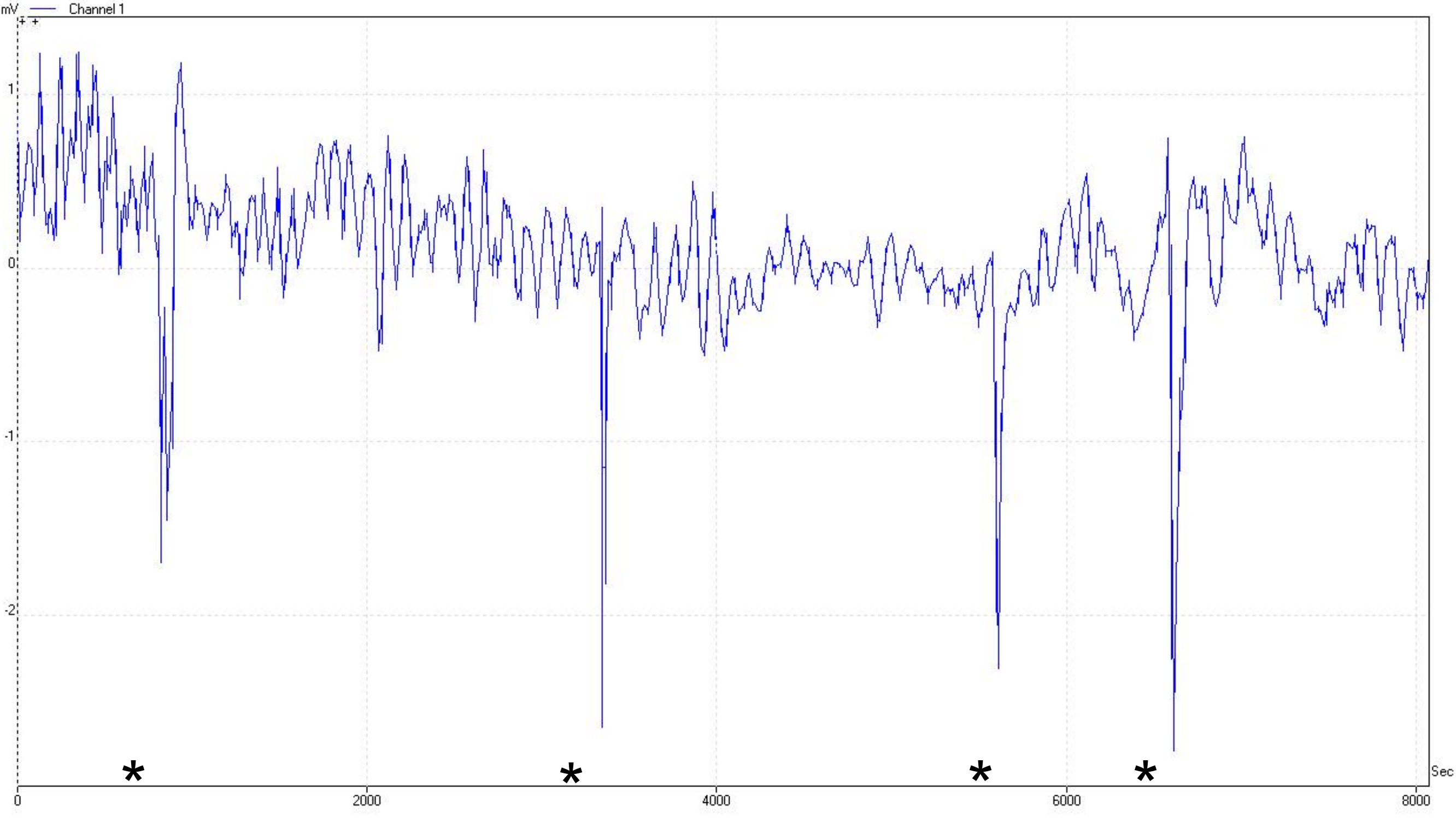}}
\caption{Typical responses of Physarum to deflection of hairs. Vertical axis is an electrical potential value in mV, horizontal axis is time in seconds.
(a)~Physarum responds with a high-amplitude impulse and envelop of four to five waves. Moment of hair deflection is shown by arrow.
(b)~hair deflected, 30 times per each stimulation, at 890~sec, 3390~sec, 5630~sec and 6680~sec, moments of stimulation are shown by asterisks.}
\label{envelopa}
\end{figure}

\begin{figure}[!tbp]
\centering
\includegraphics[width=1.1\textwidth]{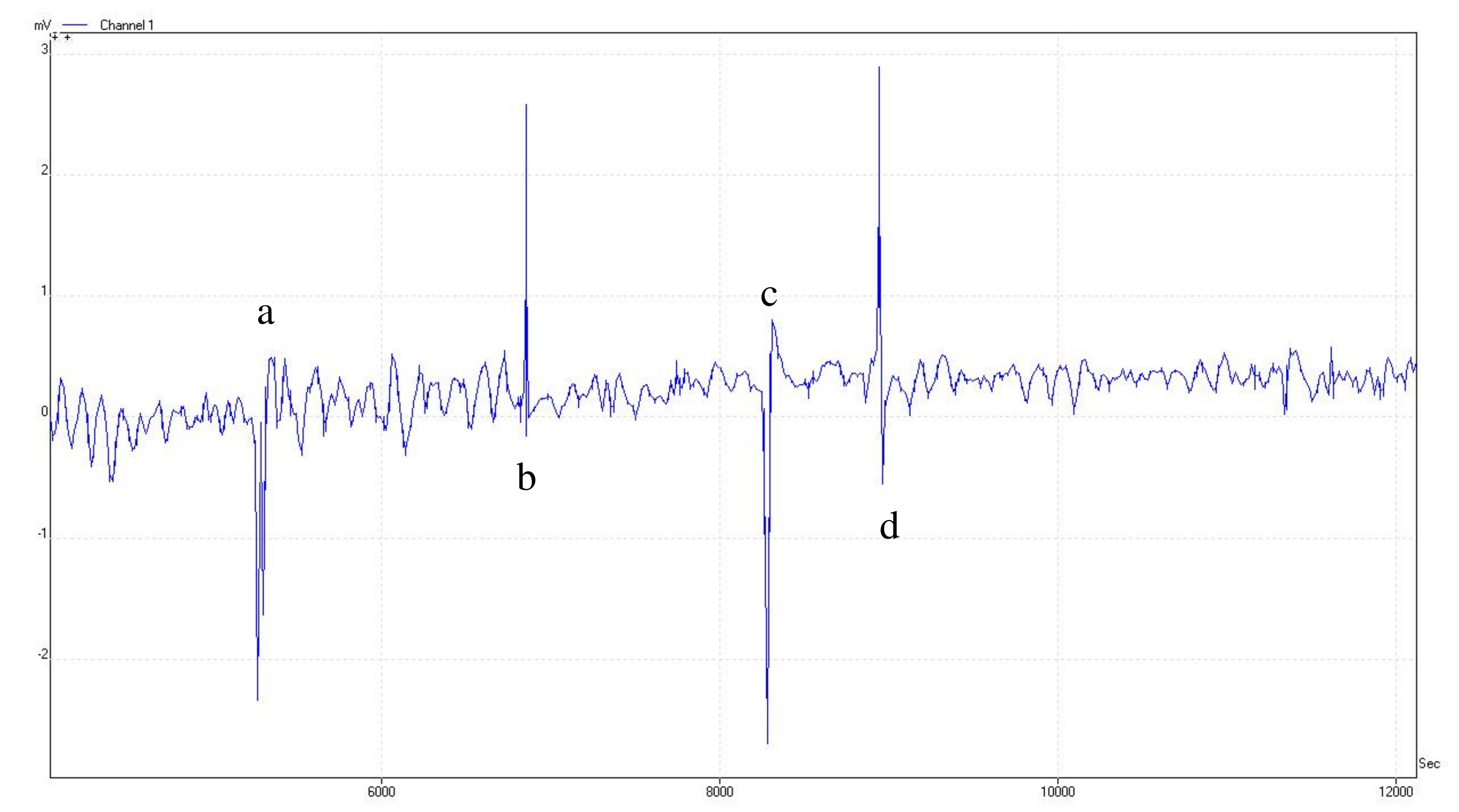}
\caption{Physarum's response deflecting hairs in turns. Vertical axis is an electrical potential value in mV, horizontal axis is time in seconds. 
(a) and (c) Hair on recording electrode is deflected 30 times. (b) and (d) Hair on reference electrode is deflected 30 times.}
\label{hairs(p)_030313}
\end{figure}

A typical response of Physarum to stimulation is shown in Fig.~\ref{parameters}. The response is comprised of an immediate response 
(Fig.~\ref{parameters}a, B): a high-amplitude impulse and a prolonged response (Fig.~\ref{parameters}a, C). See experimental recording in 
Fig.~\ref{envelopa}a. High-amplitude impulse is always well pronounced, prolonged response oscillations can sometimes be distorted by 
other factors, e.g. growing branches of a protoplasmic tube or additional strands of plasmodium propagating 
between the agar blobs. Responses are repeatable not only in different experiments but also during several rounds of stimulation in the same experiment. An example is shown in Fig.~\ref{envelopa}b. Physarum responds with a high-amplitude impulse to the first package of
stimulation, 890~sec, yet prolonged envelope response is not visible. Subsequent packages of hair deflection receives both immediate and 
well pronounced prolonged responses (Fig.~\ref{envelopa}b).  

\begin{table}[!tbp]
\caption{Statistics of Physarum response to deflection of hairs calculated in 25 experiments.}
\begin{tabular}{l|ccc}
  					& \footnotesize{Average} & \footnotesize{Stnd deviation} & \footnotesize{Median}  \\\hline
$V$						& 4.0		& 2.1		& 3.1 \\
$V'$						& 0.7		&  0.5		& 0.5 \\
$V''$						& 1.51	& 1.36	& 0.94 \\ 
$D$		 				&  135.9	& 72.0	& 129.0 \\
$D'$						& 104.7	& 25.9	& 95	\\
$D''$						& 112.9	& 17.9		& 1.21 \\
$e$						& 3.15	& 1.21	& 3	\\
\end{tabular}
\label{impulsestatistics}
\end{table}

In 25 experiments we calculated the following characteristics, see Fig.~\ref{parameters}b: amplitudes of oscillations before  $V'$ 
and after $V''$ stimulation, and of the immediate response impulse $V$, and width of impulses before $D'$ and after 
stimulation $D''$, and of immediate response $D$.   Statistics of the characteristics is shown in Tab.~\ref{impulsestatistics}.  
In our particular setup, keep in mind that signal's strength is reduced due to agar blobs and a single protoplasmic tube connecting electrodes, average 
amplitude of oscillations before stimulation is 0.7~mV and after oscillations, in the envelop of waves, is 1.51~mV. Amplitude of 
the immediate response is 4~mV in average. A prolonged response envelop has 3-4 waves.  Width of oscillation impulses becomes slightly shorter
after stimulation.  Dispersion of amplitude and width values around average values are substantial (Tab.~\ref{impulsestatistics}). This may be 
because electrical activity of Physarum and its response is determined by exact topology of a protoplasmic network wrapping agar blobs, 
and geometry of branching of the protoplasmic tube connecting the blobs.

\begin{figure}[!tbp]
\centering
\subfigure[]{\includegraphics[width=0.9\textwidth]{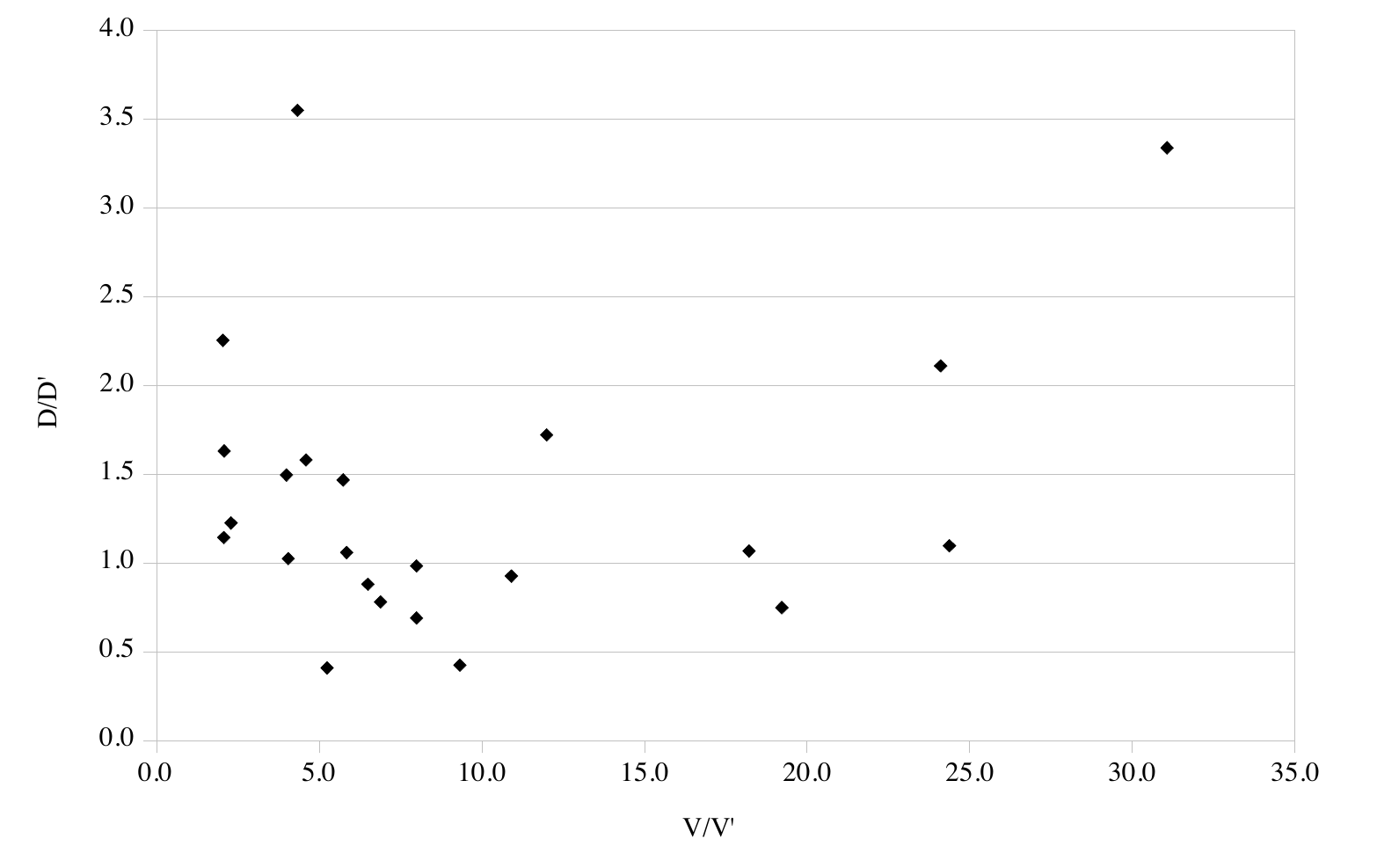}}
\subfigure[]{\includegraphics[width=0.9\textwidth]{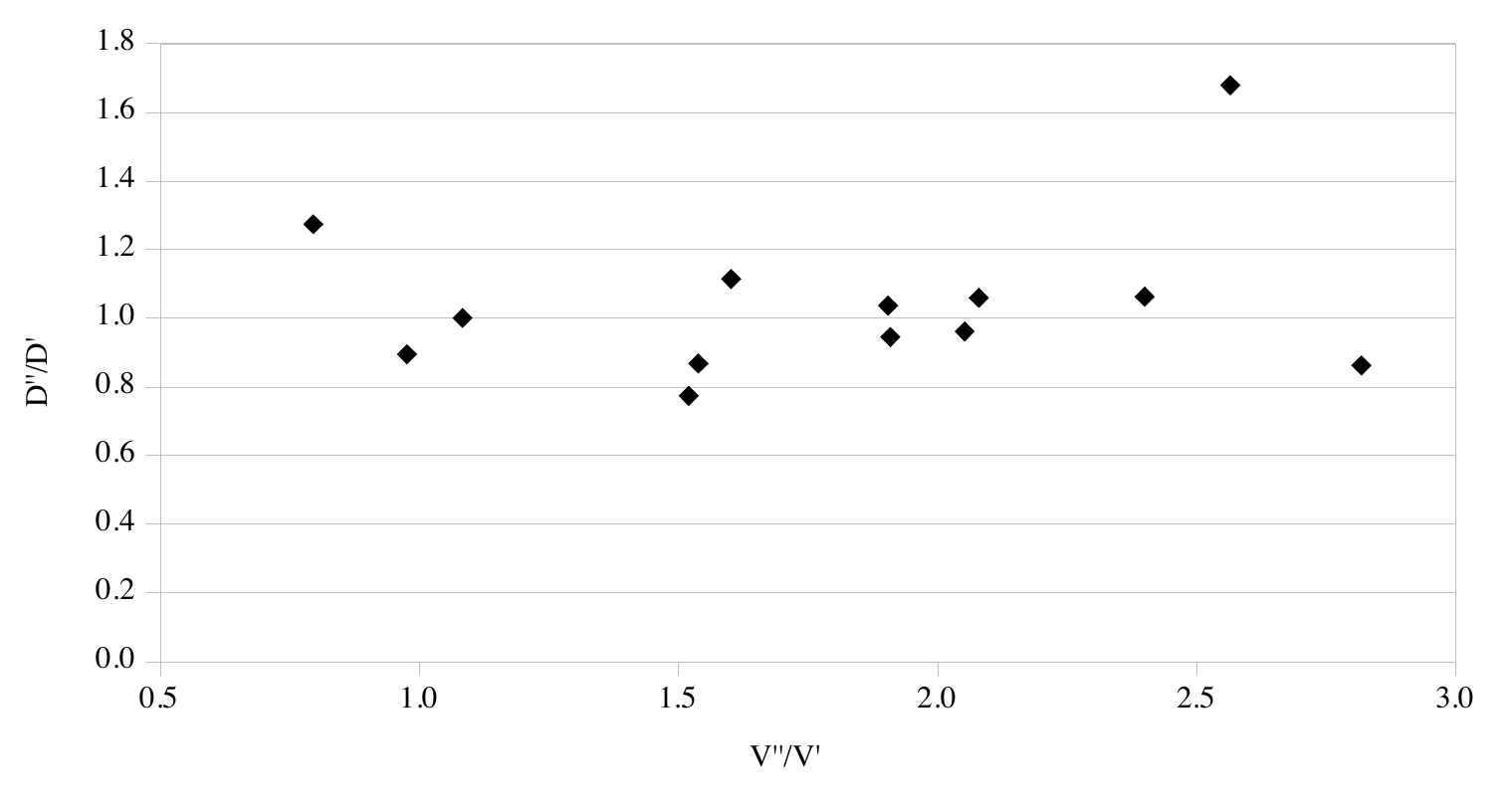}}
\caption{Maps of signal to noise ratios (SNR) of an immediate response~(a),  and prolonged response~(b) in 25 experiments. 
(a)~SNR of the immediate response is calculated as $V/V'$ and $D/D'$. 
(b)~SNR of the prolonged response is calculated as $V''/V'$ and $D''/D'$.
 }
\label{SNRmaps}
\end{figure}

Signal to noise ratio (SNR) is an important characteristic of a sensor. Physarum's electrical potential constantly oscillates. Thus we assume 
a 'noise' is a background oscillatory pattern of an undisturbed Physarum and 'signal' is an immediate response (Fig.~\ref{parameters}a, B)
and a prolonged response, an envelop of waves (Fig.~\ref{parameters}a, C). Maps of SNR obtained in laboratory experiments are 
shown in Fig.~\ref{SNRmaps}.  Experimental plots of immediate response's SNR are well grouped around average SNR of amplitude 5.7 
and SNR of width 1.29 (Fig.~\ref{SNRmaps}a).  SNR of amplitude of a prolonged response varies from 1 to almost 3 with average  2.2, 
while SNR of width is around 1 (Fig.~\ref{SNRmaps}b).  Analysis of SNRs shows that amplitudes of immediate and prolonged responses are 
robust indicators of stimulation response.

\section{Hairy balls and slimy whiskers}

\begin{figure}[!tbp]
\centering
\subfigure[]{\includegraphics[width=0.4\textwidth]{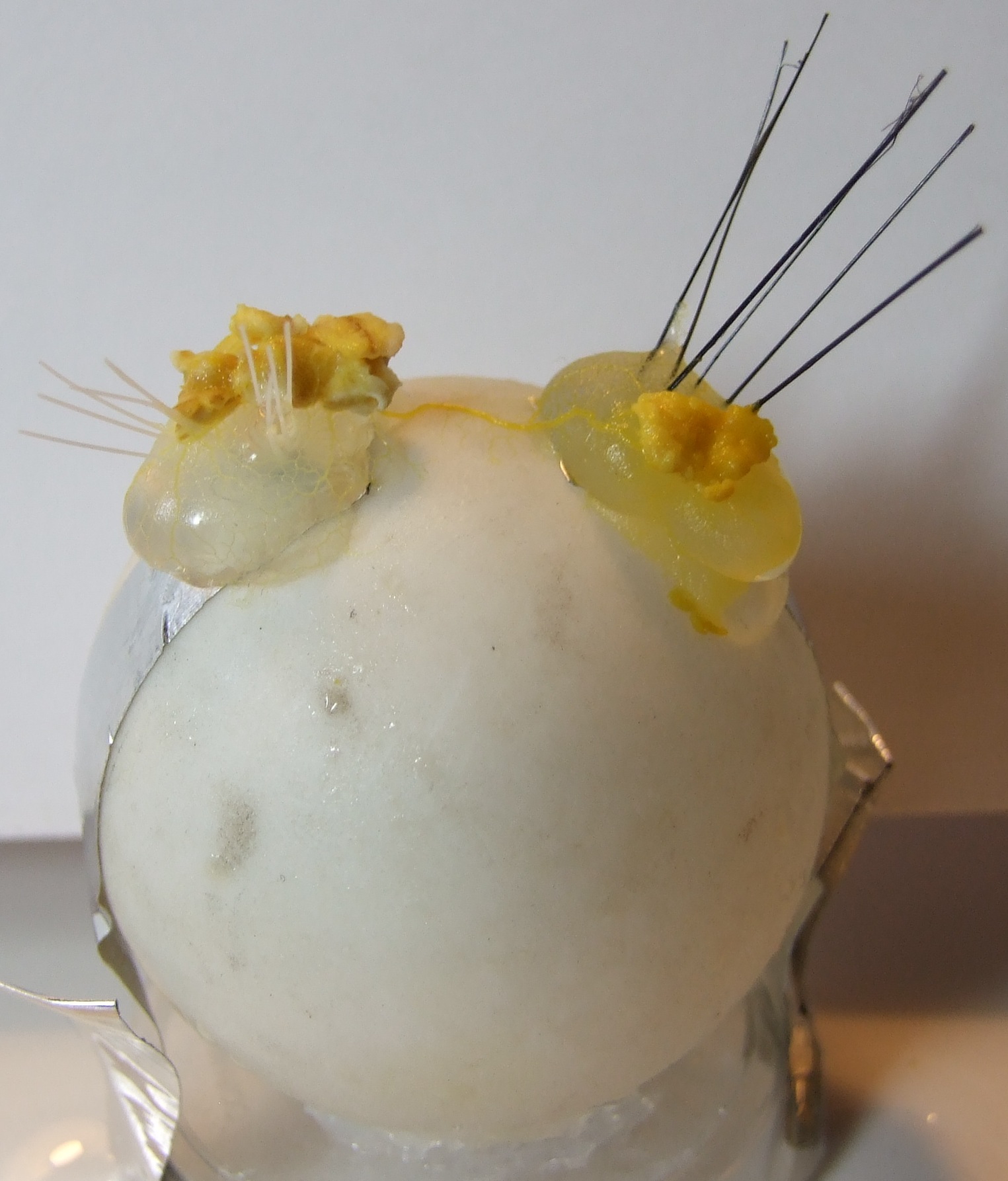}}
\subfigure[]{\includegraphics[width=0.37\textwidth]{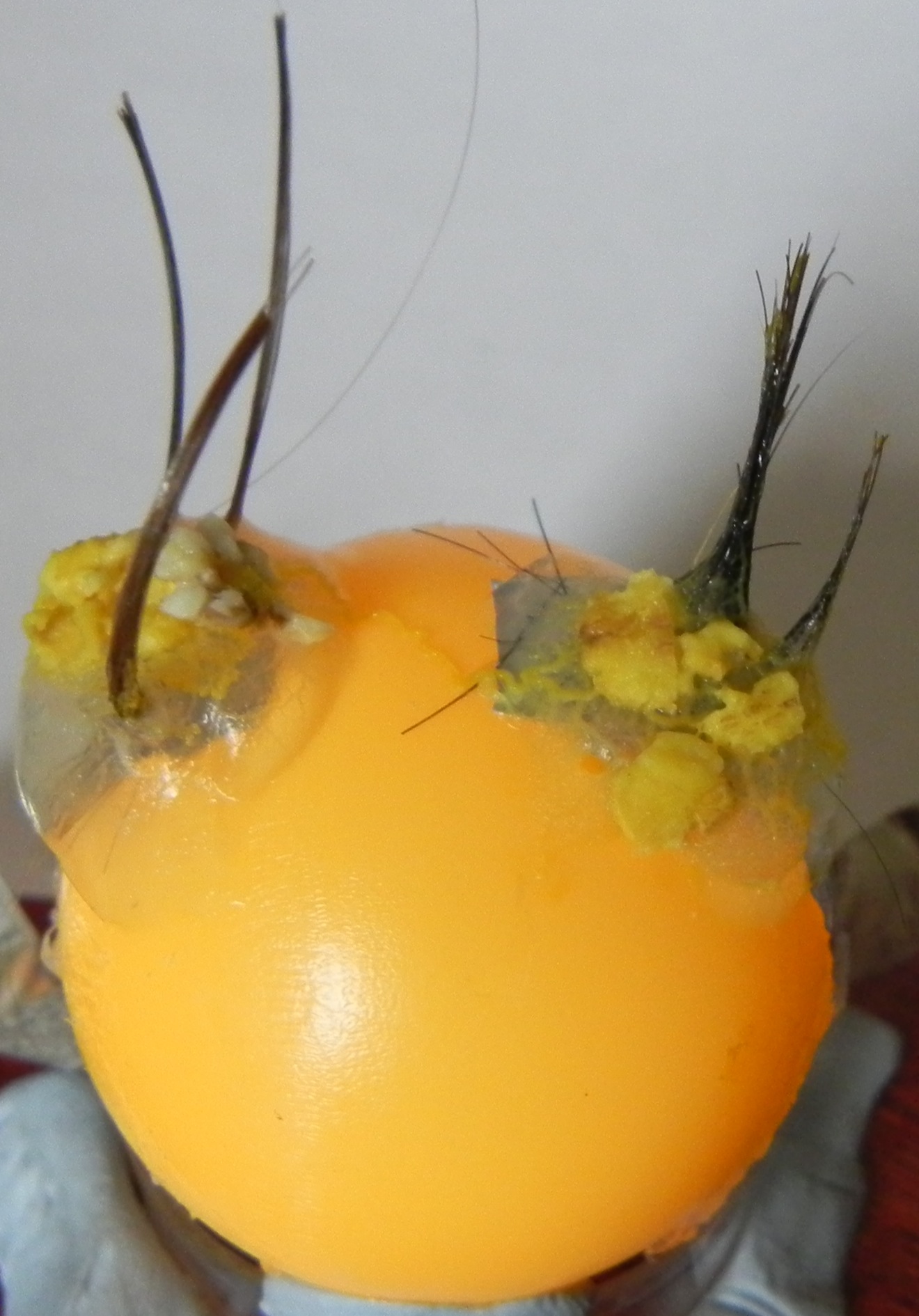}}
\caption{Physarum tactile hairs on ping-pong balls.}
\label{balls}
\end{figure}

\begin{figure}[!tbp]
\centering
\subfigure[]{\includegraphics[width=0.9\textwidth]{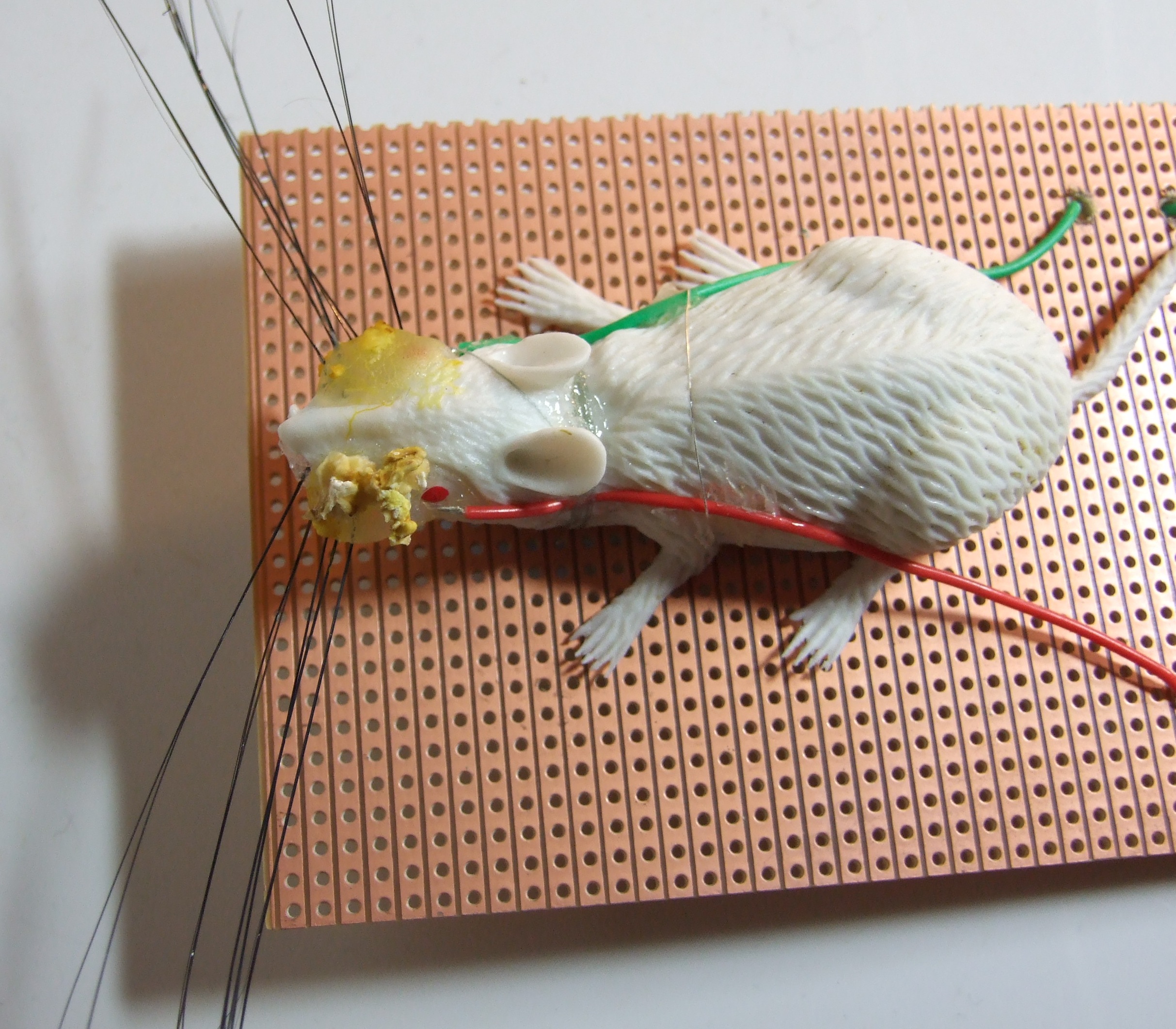}}
\subfigure[]{\includegraphics[width=0.9\textwidth]{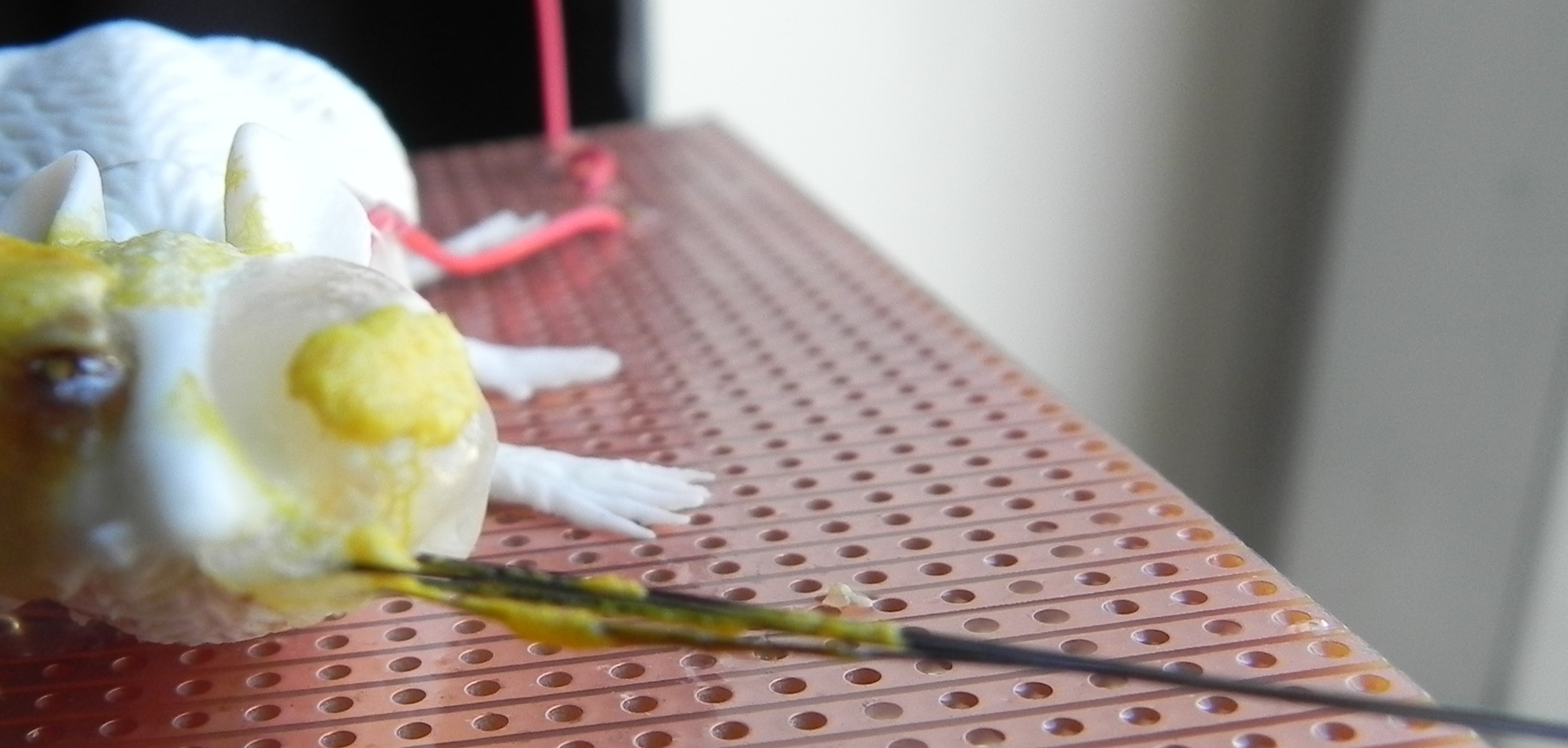}}
\caption{Rubber mouse equipped with Physarum whiskers. (a)~Top view of the setup. 
(b)~Whiskers are partly colonised by Physarum.}
\label{mouse}
\end{figure}

Will slime mould based tactile hairs work in a less ideal, than described in the previous section, environment? 
To evaluate validity of the approach we assembled Physarum tactile hair setups on  ping-pong balls (Fig.~\ref{balls}) and
a rubber mouse (Fig.~\ref{mouse}). The balls were equipped with planar aluminium foil electrodes (the same as used in original setup) and
the mouse with a hook-up wire electrodes (silver plated single core wires, cross-section area 0.23~mm$^2$, resistance 23.6 $\Omega$/1000ft).
The wire electrodes were fixed to the mouse using silicon glue (Silastic medical adhesive silicon Type A, Dow Corning). Hairs on the balls 
were made of human hairs and whiskers on the mouse of polyurethane bristles fixed to the objects' surface using Silastic silicon glue. 
We have not collected statistics but rather undertook few experiments to evaluate feasibility of the approach. 

In the ping-balls setup Physarum linked blobs of agar on electrodes with a single protoplasmic tube. 
In experiments with mouse, tubes connecting agar blobs in the base of whiskers sometimes propagated across the mouse's 
nose bridge but sometimes inframaxillary. In most experiments hairs and whiskers were partly colonised by Physarum. Degree of colonisation varied from one sevenths of a bristle to almost (Fig.~\ref{balls}) over the half of the hair/whisker length (Fig.~\ref{mouse}b)

\begin{figure}[!tbp]
\centering
\subfigure[]{\includegraphics[width=0.9\textwidth]{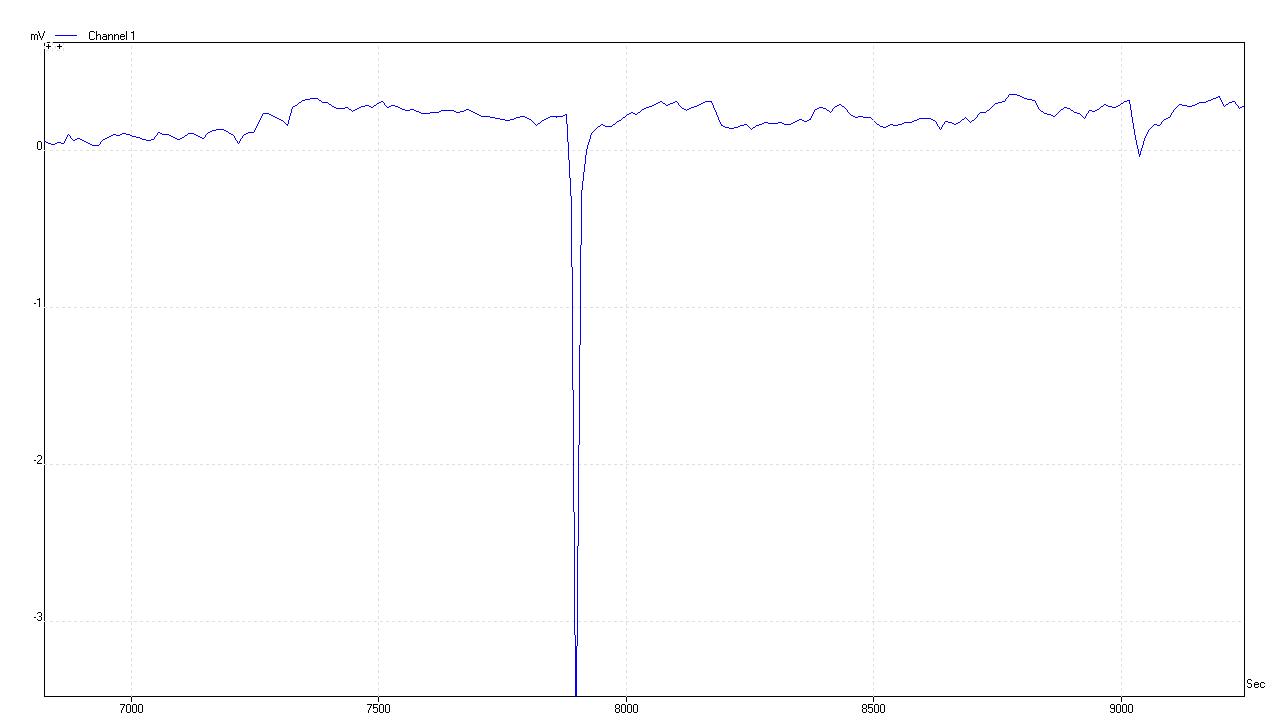}}
\subfigure[]{\includegraphics[width=0.9\textwidth]{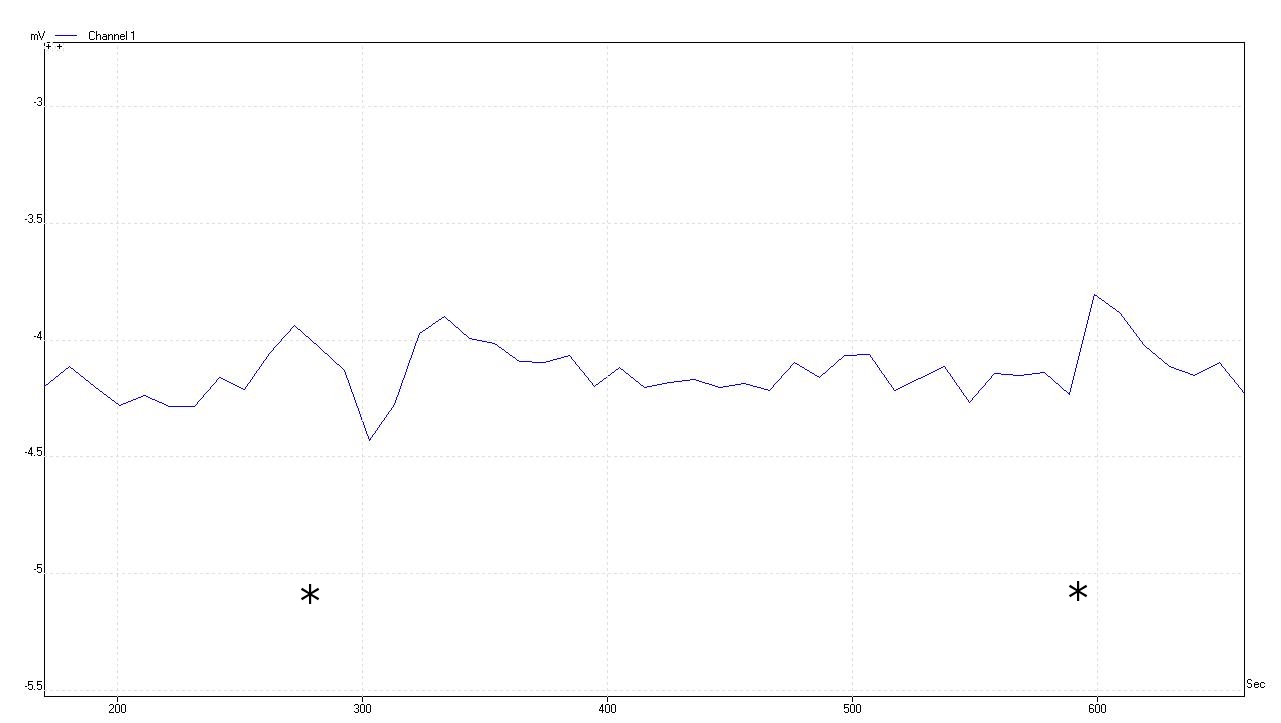}}
\caption{Physarum response to hairs deflection on a ping-pong balls (Fig.~\ref{balls}). Vertical axis is an electrical potential 
value in mV, horizontal axis is time in seconds.
(a)~Hairs on agar blob on recording electrode of  yellow ball (Fig.~\ref{balls}b) are deflected 60 times at c. 7900~sec of experiments.
(b)~Hairs on agar blob on recording electrode of white ball  (Fig.~\ref{balls}a) are deflected 60 times at c. 290~sec and 
hairs on reference electrode are deflected 60 times at c. 580~sec, both moments of hairs deflections are shown by asterisks.}
\label{ballsresponse}
\end{figure}

\begin{figure}[!tbp]
\centering
\subfigure[]{\includegraphics[width=0.9\textwidth]{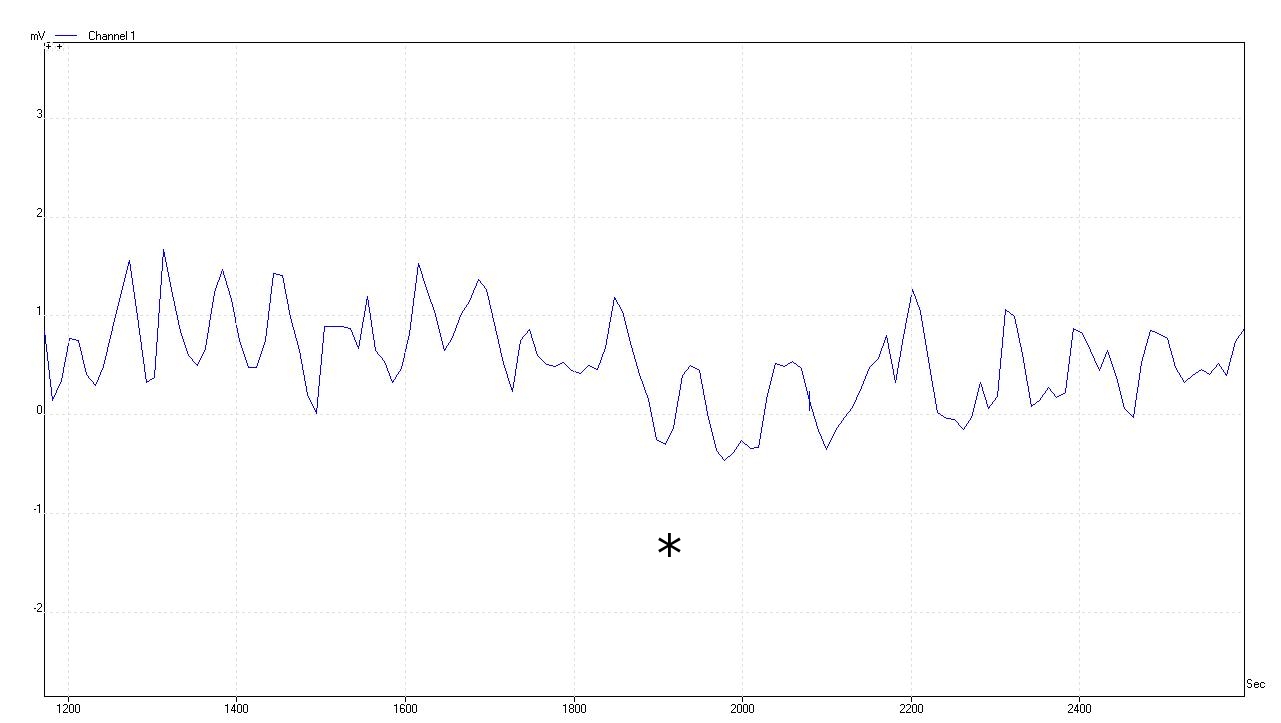}}
\subfigure[]{\includegraphics[width=0.9\textwidth]{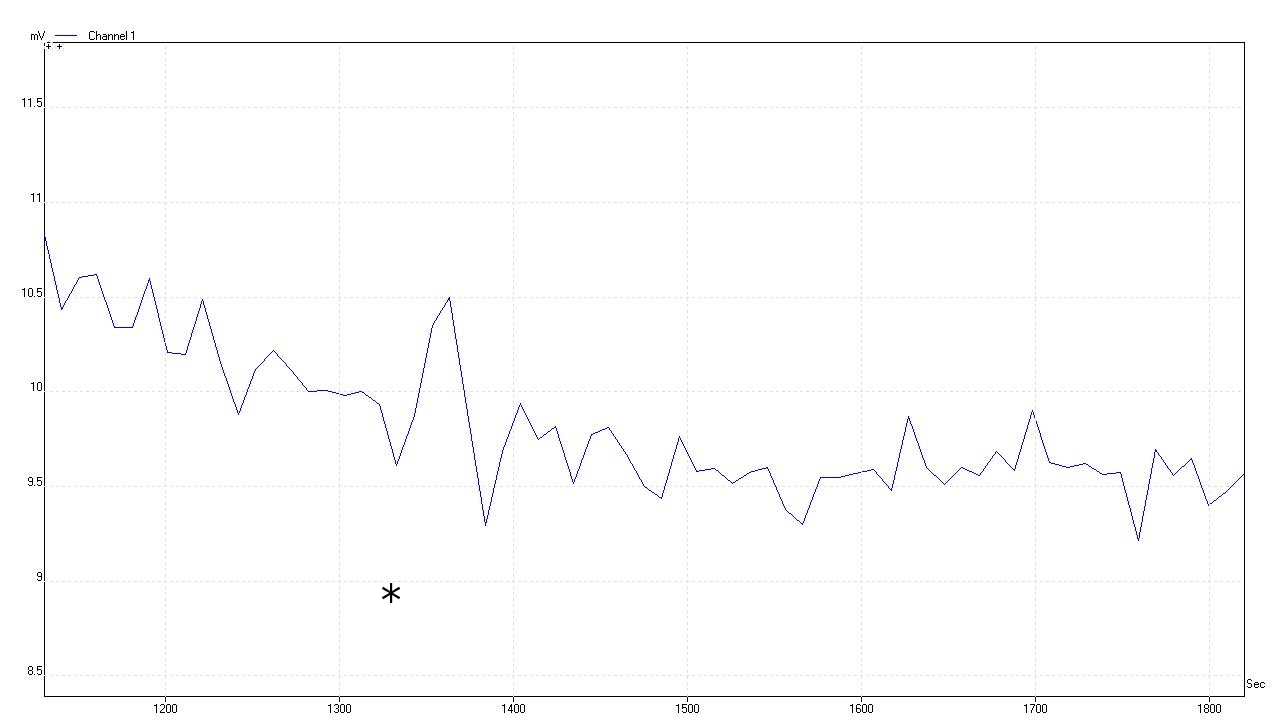}}
\caption{Physarum whiskers response on a mouse (Fig.~\ref{mouse}). Vertical axis is an electrical 
potential value in mV, horizontal axis is time in seconds.
(a)~Whiskers are deflected vertically 60 times at c.~1900~sec. 
(a)~Whiskers are deflected vertically 60 times at c.~1320~sec. 
Moments of stimulation are marked by asterisks.
}
\label{mouseresponse}
\end{figure}

Examples of responses to repeated deflections  of hairs and whiskers are shown in Figs.~\ref{ballsresponse} and \ref{mouseresponse}. 
The responses are variable. Thus, Physarum hairs on a yellow ping-pong ball (Fig.~\ref{balls}b)  reacted with an impulse of 4.3~mV to 
a repeated deflection of hairs (Fig.~\ref{ballsresponse}a).  When hairs on agar blobs on recording and reference electrodes are 
deflected in turns Physarum responds, with impulses of different signs but the same amplitude of  0.5~mV (Fig.~\ref{ballsresponse}b).
In a response illustrated in Fig.~\ref{mouseresponse}a baseline of electrical potential drops by 0.2~mV but recovers its original 
value after  270-300 seconds. A response impulse shown in Fig.~\ref{mouseresponse}b is twice of amplitude of the 
background impulses.   

SNRs of the responses illustrated are 15 in Figs.~\ref{ballsresponse}a, 2 in Figs.~\ref{ballsresponse}b, 0.9 in 
Figs.~\ref{mouseresponse}a and 2 in Figs.~\ref{mouseresponse}b. Thus in most cases response signal is well distinguishable from 
background impulses. 

\clearpage

\section{Discussion}

In laboratory experiments we designed a bio-hybrid system imitating some features of  a tactile hair of a spider.
A neuron transducing a mechanical deflection of a chair into an electrical response is physically imitated by slime mould 
\emph{Physarum polycephalum}.  Two types of responses are detected: immediate response with a high-amplitude  impulse 
and a prolonged response in a form of a wave envelop.  The slime mould tactile hairs show a reasonable value of a signal to 
noise ratio: around 6 for an immediate response and 2 for a prolonged response.  We have also installed Physarum hairs and whiskers on 
a non-planar objects and demonstrated that a signal to noise ratio of at least 2 is reached. Thus, we can claim that Physarum based 
tactile hairs and whiskers might play a significant role in future designs of bio-hybrid robotic devices. 

Our designs of slime mould based tactile hairs make an elegant addition to existing prototypes of bio-hybrid sensors incorporating 
live cells as parts of transduction system~\cite{Taniguchi_2010,Cheneler_2012} and open totally new dimension in engineering 
sensing technologies. Living substrates used for the bio-hybrid systems require  connective pathways to deliver nutrients and 
remove products of cell metabolism~\cite{Lucarotti_2013}. Some approaches to deal with this problem are based on 
micro-fabrication of vascular networks~\cite{shin_2004}. Slime mould based sensors does not require any auxiliary 'life support'
system, the cell propagates on the electrodes, consumes nutrients  from sources of food supplied and damps waste products in a 
substrate by itself.

Advantages of the proposed Physarum hairs are self-growth of sensors, low-power consumption, almost zero costs, reasonably good sensitivity, 
and a high signal to noise ratio. Disadvantages include the Physarum hairs/whiskers response dependence on a 
morphology of protoplasmic tubes and networks (which are in a state of continuous flux), susceptibility to temperature, light and humidity change, and 
relatively short period of functionality (usually 3-4 days).

The devices described  in the paper are instances of \emph{wetware of a secondary class of living technologies}~\cite{bedau_2010}. 
Rephrasing, Bedau et al.~ \cite{bedau_2010}, we can say that Physarum tactile hairs proposed are artificial because they are created by our intentional activities yet they are totally natural because they grow and respond to environmental stimuli by their own biological laws. 

Further research in the slime mould based tactile hairs will aim to answer the following questions. What are biophysical 
mechanisms of hair/whisker to Physarum to electrical output transduction?  Do cellular membrane and outer wall of 
Physarum's protoplasmic tubes react to stretching and twisting in same manner as e.g. bacteria do~\cite{hamil_2001}? 
How does a response's strength (in amplitude and duration) depends on a number of hairs/whiskers deflected? 
How to keep the 'slimy whiskers' functioning for weeks and months? What is an adaptability pattern of the Physarum hairs
when they are exposes to repeated rounds of stimulation?

\end{document}